\documentclass[seceq,supplement]{ptptex}
\title{Representation of the nonequilibrium steady-state distribution function
for sheared granular systems}

\author{
Song-Ho \textsc{Chong}$^1$,
Michio \textsc{Otsuki}$^2$ and 
Hisao \textsc{Hayakawa}$^3$}

\inst{
$^1$Institute for Molecular Science, Okazaki 444-8585, Japan\\
$^2$Department of Physics and Mathematics, Aoyama Gakuin University,\\
Kanagawa 229-8558, Japan\\
$^3$Yukawa Institute for Theoretical Physics, Kyoto University,\\
Kyoto 606-8502, Japan}

\abst{
We derive the representation of the nonequilibrium steady-state distribution function
which is expressed in terms of the excess free energy production.
This representation resembles the one derived recently by
Komatsu and Nakagawa [Phys.~Rev.~Lett.~{\bf 100} (2008), 030601] 
resting on the use of microscopic time-reversal symmetry, but 
our representation applies also to sheared granular 
systems in which such a symmetry is broken.} 

\newcommand{\bv}[1]{\boldsymbol #1}

\begin{document}

\maketitle

\section{Introduction}

Developing statistical mechanics for nonequilibrium steady states is one of the
most challenging problems in theoretical physics~\cite{Zubarev74,McLennan88,Sasa06,Evans08}.
Among the most remarkable outcomes from those studies devoted to such a development 
have been the generalized Green-Kubo relation~\cite{Morriss87}
and various forms of work and fluctuation theorems~\cite{FT-papers,Evans02}.
A suggestive representation of the nonequilibrium
steady-state distribution function in terms of the excess entropy
production has recently been developed~\cite{Komatsu08,Komatsu09}, and
its connection to the steady-state thermodynamics
has been discussed~\cite{Komatsu08b}.

However, most of these studies rest on the use of microscopic
time reversality or the local detailed balance, and the outcomes therefrom
cannot be applied literally to 
macroscopic, irreversible dissipative systems like granular fluids~\cite{Puglisi05} 
despite manifest similarities~\cite{Feitosa04}.
Recently, we have demonstrated that 
the generalized Green-Kubo relation and the
integral form of the fluctuation theorem can be derived 
without assuming the microscopic time reversality~\cite{Chong09b},
thus extending the major achievements so far in the nonequilibrium statistical mechanics 
to irreversible dissipative systems.
A purpose of the present paper is to further continue this line 
of work, and to explore the representation of the steady-state distribution function
which resembles the one derived in Refs.~\citen{Komatsu08} and \citen{Komatsu09},
but applies also to irreversible dissipative systems.
This will be exemplified for uniformly sheared granular systems.

The paper is organized as follows.
In Sec.~\ref{sec:exact-starting-points}, 
after introducing basic equations of motion for uniformly sheared granular systems,
we summarize exact relations which serve a basis in our
subsequent formulation.
We then derive in Sec.~\ref{sec:cumulant-expansion-representation}
the representation of the steady-state distribution function
which resembles the one developed in Refs.~\citen{Komatsu08} and \citen{Komatsu09},
but is applicable also to systems in which the microscopic time-reversal
symmetry is broken.
The paper is summarized in Sec.~\ref{sec:summary}. 
Appendix~\ref{appendix:B} is devoted to a technical derivation of the result
which is used in the main text.

\section{Exact starting points}
\label{sec:exact-starting-points}

In this section, we introduce basic equations of motion
for sheared granular systems and corresponding Liouville equations. 
We then summarize some exact relations which serve a basis in our subsequent formulation.

\subsection{SLLOD equations}

The system of our interest consists of $N$ smooth granular particles of mass $m$ which is immersed 
in a volume $V$ and is subjected to stationary shearing characterized by the
shear-rate tensor $\mbox{\boldmath $\kappa$}$.
We assume that each granular particle is a soft-shere, and the contact force
acts only on the normal direction. 
For a simple uniform shear with velocity
along the $x$-axis and its gradient along the $y$-axis, which we consider
throughout this paper, the shear-rate tensor is 
$\kappa_{\alpha \beta} = \dot{\gamma} \delta_{\alpha x} \delta_{\beta y}$
with $\dot{\gamma}$ denoting the shear rate.
Under homogeneous shear, a linear streaming velocity profile
$\bv{\kappa} \cdot \bv{r}$ is induced at position $\bv{r}$.
Newtonian equations of motion describing such a homogeneously sheared system
are the SLLOD equations~\cite{Evans08}
\begin{subequations}
\label{eq:SLLOD}
\begin{eqnarray}
\dot{\bv{r}}_{i} &=& \frac{\bv{p}_{i}}{m} + \mbox{\boldmath $\kappa$} \cdot \bv{r}_{i},
\label{eq:SLLOD-a}
\\
\dot{\bv{p}}_{i} &=& \bv{F}_{i}^{\rm (el)} + \bv{F}_{i}^{\rm (vis)} - \mbox{\boldmath $\kappa$} \cdot \bv{p}_{i}.
\label{eq:SLLOD-b}
\end{eqnarray}
\end{subequations}
Here $\bv{r}_{i}$ refers to the position of the $i$th particle, and 
$\bv{p}_{i}$ is the peculiar, or thermal, momentum 
defined with respect to the streaming velocity 
$\bv{\kappa} \cdot \bv{r}_{i}$.
The conservative force $\bv{F}_{i}^{\rm (el)} = - \partial U / \partial \bv{r}_{i}$ 
with the total interaction potential $U$ is 
given by a sum $\bv{F}_{i}^{\rm (el)} = \sum_{j \ne i} \bv{F}_{ij}^{\rm (el)}$
of the elastic repulsive forces exerted on the $i$th particle by others
\begin{equation}
\bv{F}_{ij}^{\rm (el)} = - \frac{\partial u(r_{ij})}{\partial \bv{r}_{ij}} =
\Theta(d-r_{ij}) f(d-r_{ij}) \hat{\bv{r}}_{ij},
\label{eq:F-el-def}
\end{equation}
where $u(r)$ is the pair potential; 
$\bv{r}_{ij} = \bv{r}_{i} - \bv{r}_{j}$, $r_{ij} = | \bv{r}_{ij} |$, $\hat{\bv{r}}_{ij} = \bv{r}_{ij} / r_{ij}$;
$d$ denotes the particle diameter; 
$\Theta(x)$ is the Heaviside step function.
Typical forms of the elastic repulsive force are
$f(x) \propto x$ (linear model) and $f(x) \propto x^{3/2}$ (Hertzian model).
Similarly, the viscous dissipative force $\bv{F}_{i}^{\rm (vis)}$
due to inelastic collisions between particles is represented by 
a sum $\bv{F}_{i}^{\rm (vis)} = \sum_{j \ne i} \bv{F}_{ij}^{\rm (vis)}$ of
two-body contact forces 
\begin{equation}
\bv{F}_{ij}^{\rm (vis)} = - \Theta(d-r_{ij}) \zeta(d-r_{ij}) (\bv{g}_{ij} \cdot \hat{\bv{r}}_{ij}) \hat{\bv{r}}_{ij}.
\label{eq:F-vis-def}
\end{equation}
It is proportional to the relative velocity $\bv{g}_{ij} \equiv \dot{\bv{r}}_{i} - \dot{\bv{r}}_{j} =
(\bv{p}_{i} - \bv{p}_{j})/m + \bv{\kappa} \cdot \bv{r}_{ij}$
of colliding particles, and for this reason,
the SLLOD equations (\ref{eq:SLLOD}) are not
invariant under the time-reversal map 
$\{ \bv{r}_{i}, \bv{p}_{i}, \dot{\gamma} \} \to \{ \bv{r}_{i}, -\bv{p}_{i}, -\dot{\gamma} \}$.
The amount of energy dissipation upon inelastic collisions 
is characterized by the viscous function $\zeta(x)$ which is typically assumed to be
constant or modeled as $\zeta(x) \propto x^{1/2}$.

\subsection{The Liouville equation}

For nonequilibrium systems, 
the form of the Liouville equation commonly used 
for Hamiltonian systems should be properly generalized to take into account
the effect of phase-space compression~\cite{Evans08}.
The Liouville equation for phase variables, say $A(\bv{\Gamma})$,
where $\bv{\Gamma} = (\bv{r}^{N}, \bv{p}^{N})$ stands for a phase-space point,
is given by
\begin{equation}
\frac{d}{dt} A(\bv{\Gamma}) =
\dot{\bv{\Gamma}} \cdot 
\frac{\partial}{\partial \bv{\Gamma}}
A(\bv{\Gamma}) \equiv
i {\cal L}(\bv{\Gamma}) A(\bv{\Gamma}).
\label{eq:Lp}
\end{equation}
The operator $i{\cal L}(\bv{\Gamma})$ is referred to as the
$p$-Liouvillean.
The explicit expression for $i{\cal L}(\bv{\Gamma})$ corresponding to the
SLLOD equations (\ref{eq:SLLOD}) shall be
written for later convenience in the form
\begin{subequations}
\label{eq:explicit-iLp}
\begin{equation}
i {\cal L}(\bv{\Gamma}) = i {\cal L}_{0}(\bv{\Gamma}) + i \Delta {\cal L}(\bv{\Gamma})
\label{eq:explicit-iLp-a}
\end{equation}
with 
\begin{eqnarray}
i{\cal L}_{0} (\bv{\Gamma}) &=& 
\sum_{i} 
\Bigl[ 
  \frac{\bv{p}_{i}}{m} \cdot \frac{\partial}{\partial \bv{r}_{i}} + \bv{F}_{i}^{\rm (el)} \cdot \frac{\partial}{\partial \bv{p}_{i}} 
\Bigr],
\label{eq:explicit-iLp-b}
\\
i \Delta {\cal L}(\bv{\Gamma}) &=&
\sum_{i} \Bigl[ 
  (\mbox{\boldmath $\kappa$} \cdot \bv{r}_{i}) \cdot \frac{\partial}{\partial \bv{r}_{i}} -
  (\mbox{\boldmath $\kappa$} \cdot \bv{p}_{i}) \cdot \frac{\partial}{\partial \bv{p}_{i}}
\Bigr] +
\sum_{i} \bv{F}_{i}^{\rm (vis)} \cdot \frac{\partial}{\partial \bv{p}_{i}} .
\label{eq:explicit-iLp-c}
\end{eqnarray}
\end{subequations}
Here, $i{\cal L}_{0}(\bv{\Gamma})$ represents an ``unperturbed" part which is derivable from the Hamiltonian describing
the internal energy of the system
\begin{equation}
H(\bv{\Gamma}) = \sum_{i} \frac{\bv{p}_{i}^{2}}{2m} + U(\bv{r}^{N}),
\label{eq:H-def}
\end{equation}
while the rest $i \Delta {\cal L}(\bv{\Gamma})$ is due to the driving forces towards the nonequilibrium states.
Here, not only the shearing force, but also the viscous dissipative force should be considered
as the driving force. 
This notion is clear for the sheared granular system under study: 
even when the shearing force is absent, 
the system gets of out equilibrium due to the dissipative force
describing inelastic collisions between particles.

The formal solution to the Liouville equation (\ref{eq:Lp}) can be written
in terms of the
$p$-propagator $\exp[ i {\cal L}(\bv{\Gamma}) t]$ as
\begin{equation}
A(\bv{\Gamma}(t)) = \exp[i {\cal L}(\bv{\Gamma}) t] \, A(\bv{\Gamma}).
\label{eq:p-propagator}
\end{equation}
Hereafter, the absence of the argument $t$ implies that
associated quantities are evaluated at $t=0$,
and the dependence on $\bv{\Gamma}$ shall often be dropped 
for brevity like $A(t) = A(\bv{\Gamma}(t))$ and $i {\cal L} = i {\cal L}(\bv{\Gamma})$.

On the other hand, 
the Liouville equation for the 
phase-space distribution function $f(\bv{\Gamma},t)$,
normalized such that $\int d\bv{\Gamma} \, f(\bv{\Gamma},t) = 1$, 
is given by~\cite{Evans08}
\begin{equation}
\frac{\partial f(\bv{\Gamma},t)}{\partial t} =
- \frac{\partial}{\partial \bv{\Gamma}} \cdot \Bigl[ \dot{\bv{\Gamma}} f(\bv{\Gamma},t) \Bigr] =
- \Bigl[ \dot{\bv{\Gamma}} \cdot 
       \frac{\partial}{\partial \bv{\Gamma}} +
       \Lambda(\bv{\Gamma}) 
\Bigr] f(\bv{\Gamma},t) \equiv
- i {\cal L}^{\dagger}(\bv{\Gamma}) f(\bv{\Gamma},t).
\label{eq:Lf}
\end{equation}
The operator $i{\cal L}^{\dagger}$ is called the
$f$-Liouvillean, and 
$\Lambda(\bv{\Gamma}) \equiv (\partial / \partial \bv{\Gamma}) \cdot \dot{\bv{\Gamma}}$ 
is referred to as the phase-space compression factor.
For the SLLOD equations (\ref{eq:SLLOD}), one has
\begin{equation}
\Lambda(\bv{\Gamma}) = \sum_{i} \frac{\partial}{\partial \bv{p}_{i}} \cdot \bv{F}_{i}^{\rm (vis)} =
- \frac{1}{m} \sum_{i} \sum_{j \ne i} \Theta(d-r_{ij}) \zeta(d-r_{ij}).
\label{eq:SLLOD-Lambda}
\end{equation}

It follows from (\ref{eq:Lp}) and (\ref{eq:Lf}) that
\begin{equation}
i{\cal L}^{\dagger}(\bv{\Gamma}) = 
i{\cal L}(\bv{\Gamma}) + \Lambda(\bv{\Gamma}).
\label{eq:relation-Liouville-operators}
\end{equation}
Thus, the Liouville operator is non-Hermitian in the presence of the
viscous dissipative force. 
Noticing from (\ref{eq:SLLOD-Lambda}) that $\Lambda(\bv{\Gamma})$ is completely determined by the 
viscous dissipative force,
$i {\cal L}^{\dagger}$ shall also be decomposed in the form of (\ref{eq:explicit-iLp}) as
\begin{subequations}
\label{eq:explicit-iLf}
\begin{equation}
i{\cal L}^{\dagger}(\bv{\Gamma}) = i {\cal L}_{0}^{\dagger}(\bv{\Gamma}) + i \Delta {\cal L}^{\dagger}(\bv{\Gamma})
\label{eq:explicit-iLf-a}
\end{equation}
with
\begin{equation}
i {\cal L}_{0}^{\dagger}(\bv{\Gamma}) = i {\cal L}_{0}(\bv{\Gamma}), \quad 
i \Delta {\cal L}^{\dagger}(\bv{\Gamma}) = i \Delta {\cal L}(\bv{\Gamma}) + \Lambda(\bv{\Gamma}).
\label{eq:explicit-iLf-b}
\end{equation}
\end{subequations}

The formal solution to the Liouville equation (\ref{eq:Lf}) reads
\begin{equation}
f(\bv{\Gamma},t) = \exp( - i {\cal L}^{\dagger} t) f(\bv{\Gamma},0), 
\label{eq:f-propagator}
\end{equation}
where $\exp( - i {\cal L}^{\dagger} t)$ is called the $f$-propagator.
For later convenience, the initial distribution function $f(\bv{\Gamma},0)$ shall be
chosen as the equilibrium one generated by $i {\cal L}_{0}^{\dagger}$, i.e.,
as the solution of 
\begin{equation}
i {\cal L}_{0}^{\dagger}(\bv{\Gamma}) f(\bv{\Gamma},0) = 0.
\label{eq:property-initial-distribution}
\end{equation}
It follows from $i{\cal L}_{0}^{\dagger} = i{\cal L}_{0}$ and (\ref{eq:explicit-iLp-b}) that
$f(\bv{\Gamma},0)$ for the present system is given by the canonical distribution function
of the inverse temperature $\beta$,
\begin{equation}
f(\bv{\Gamma},0) = \frac{1}{{\cal Z}(\beta)} e^{-\beta H(\bv{\Gamma})} \,\,\,
\mbox{with} \,\,\,
{\cal Z}(\beta) \equiv \int d\bv{\Gamma} \, e^{-\beta H(\bv{\Gamma})}.
\label{eq:canonical-distribution}
\end{equation}
As demonstrated in Ref.~\citen{Chong09b}, nonequilibrium steady-state properties
based on the canonical initial distribution function do not depend on the 
choice of $\beta$. 

Let us summarize here for later use 
some relations between $f$- and $p$-Liouvilleans and 
corresponding propagators. 
When $i{\cal L}^{\dagger}(\bv{\Gamma})$ acts on the product of the initial
distribution function $f(\bv{\Gamma},0)$ and a phase variable $A(\bv{\Gamma})$,
there holds
\begin{eqnarray}
i {\cal L}^{\dagger}(\bv{\Gamma}) [ f(\bv{\Gamma},0) A(\bv{\Gamma}) ] &=&
\frac{\partial}{\partial \bv{\Gamma}} \cdot 
[ \dot{\bv{\Gamma}} f A ] =
A \frac{\partial}{\partial \bv{\Gamma}} \cdot [ \dot{\bv{\Gamma}} f ] +
f \dot{\bv{\Gamma}} \cdot \frac{\partial}{\partial \bv{\Gamma}} A
\nonumber \\
&=&
A(\bv{\Gamma}) [ i {\cal L}^{\dagger}(\bv{\Gamma}) f(\bv{\Gamma},0) ] + 
f(\bv{\Gamma},0) [ i {\cal L}(\bv{\Gamma}) A(\bv{\Gamma})].
\label{eq:iL-dagger-fA}
\end{eqnarray}
One can show that
$i{\cal L}$ and $i{\cal L}^{\dagger}$
are adjoint operators:
\begin{equation}
\int d\bv{\Gamma} \,
[ i {\cal L}(\bv{\Gamma}) A(\bv{\Gamma}) ] \, B(\bv{\Gamma}) =
- \int d\bv{\Gamma} \,
A(\bv{\Gamma}) \,
[ i {\cal L}^{\dagger}(\bv{\Gamma}) B(\bv{\Gamma})].
\label{eq:adjoint-property}
\end{equation}
This property can be proved from the integration by parts. 
By a repeated use of this property, the following
relation for the propagators can be derived:
\begin{equation}
\int d\bv{\Gamma} \,
[ e^{i {\cal L}(\bv{\Gamma}) t} A(\bv{\Gamma})] \, B(\bv{\Gamma}) =
\int d\bv{\Gamma} \,
A(\bv{\Gamma}) \, [ e^{- i {\cal L}^{\dagger}(\bv{\Gamma}) t} B(\bv{\Gamma})].
\label{eq:unrolling}
\end{equation} 

\subsection{Generalized Green-Kubo relation}
\label{subsec:generalized-Green-Kubo-relation}

With the identity
\begin{equation}
e^{-i {\cal L}^{\dagger}(\bv{\Gamma}) t} = 1 + \int_{0}^{t} ds \, e^{-i{\cal L}^{\dagger}(\bv{\Gamma}) s} 
[-i{\cal L}^{\dagger}(\bv{\Gamma})],
\end{equation}
the formal solution (\ref{eq:f-propagator}) can be expressed as
\begin{eqnarray}
f(\bv{\Gamma},t) &=& f(\bv{\Gamma},0) +
\int_{0}^{t} ds \, e^{-i{\cal L}^{\dagger}(\bv{\Gamma}) s} [-i{\cal L}^{\dagger}(\bv{\Gamma}) f(\bv{\Gamma},0)]
\nonumber \\
&\equiv&
f(\bv{\Gamma},0) +
\int_{0}^{t} ds \, e^{-i{\cal L}^{\dagger}(\bv{\Gamma}) s} [ f(\bv{\Gamma},0) \Omega(\bv{\Gamma})],
\label{eq:f-Gamma-t}
\end{eqnarray}
where in the final equality we have defined the dissipation function $\Omega(\bv{\Gamma})$.
From an explicit calculation, one finds for the present system~\cite{Chong09b}
\begin{subequations}
\label{eq:Omega-def}
\begin{equation}
\Omega(\bv{\Gamma}) = - \beta \dot{\gamma} \sigma_{xy}(\bv{\Gamma}) - 2 \beta {\cal R}(\bv{\Gamma}) - \Lambda(\bv{\Gamma}).
\label{eq:Omega-def-a}
\end{equation}
Here, the shear-stress tensor $\sigma_{\alpha \beta}(\bv{\Gamma})$ and Rayleigh's dissipation function ${\cal R}(\bv{\Gamma})$
are respectively given by
\begin{eqnarray}
& &
\sigma_{\alpha \beta}(\bv{\Gamma}) =
\sum_{i} \Bigl[ \frac{p_{i,\alpha} p_{i,\beta}}{m} + r_{i,\alpha} (F_{i,\beta}^{\rm (el)} + F_{i,\beta}^{\rm (vis)}) \Bigr],
\label{eq:Omega-def-b}
\\
& &
{\cal R}(\bv{\Gamma}) = \frac{1}{4} \sum_{i} \sum_{j \ne i} 
\Theta(d-r_{ij}) \zeta(d-r_{ij}) (\bv{g}_{ij} \cdot \hat{\bv{r}}_{ij})^{2}.
\label{eq:Omega-def-c}
\end{eqnarray}
\end{subequations}

Let us summarize here some important properties concerning the dissipation function $\Omega(\bv{\Gamma})$.
First, one can show that the equilibrium average of the dissipation function is zero~\cite{Chong09b}:
\begin{equation}
\langle \Omega(\bv{\Gamma}) \rangle = 0.
\label{eq:Omega-0}
\end{equation}
Here and in the following, the ensemble average $\langle \cdots \rangle$ is defined over the
initial equilibrium distribution function $f(\bv{\Gamma},0)$, i.e.,
\begin{equation}
\langle \cdots \rangle \equiv \int d\bv{\Gamma} \, f(\bv{\Gamma},0) \cdots.
\label{eq:average-def}
\end{equation}
Second, there holds~\cite{Chong09b}
\begin{equation}
d\bv{\Gamma}(t) \, f(\bv{\Gamma}(t),0) = d\bv{\Gamma} \, f(\bv{\Gamma},0) 
e^{-\int_{0}^{t} ds \, \Omega(\bv{\Gamma}(s))}.
\label{eq:Omega-property-2}
\end{equation}
This relation is a consequence of the conservation of the number
of ensemble members within a comoving phase volume~\cite{Evans02}.
Finally, we notice that, if we define the time-dependent free energy via 
$\beta {\cal F}(t) \equiv \beta \langle H(t) \rangle - S(t)$ in terms of the
Gibbs entropy $S(t) \equiv - \int d\bv{\Gamma} \, f(\bv{\Gamma},t) \log f(\bv{\Gamma},t)$,
one finds
$\beta \dot{\cal F}(t) = \langle \Omega(t) \rangle$ since 
$\dot{H} = -\dot{\gamma} \sigma_{xy} - 2 {\cal R}$ and
$\dot{S}(t) = \langle \Lambda(t) \rangle$.
Thus, the average dissipation function divided by $\beta$ can be identified as
the free energy production rate.

The nonequilibrium ensemble average 
$\langle A(t) \rangle$ shall be defined via
\begin{equation}
\langle A(t) \rangle \equiv
\int d\bv{\Gamma} \,
f(\bv{\Gamma},0) \, A(t) = 
\int d\bv{\Gamma} \,
f(\bv{\Gamma},t) \, A(0).
\label{eq:neq-average}
\end{equation}
The two representations in terms of
$f(\bv{\Gamma},0)$ or 
$f(\bv{\Gamma},t)$ are equivalent
because of the relation~(\ref{eq:unrolling}).
Substituting (\ref{eq:f-Gamma-t}) for $f(\bv{\Gamma},t)$,
one obtains 
\begin{equation}
\langle A(t) \rangle = \langle A(0) \rangle +
\int_{0}^{t} ds \, \langle A(s) \Omega(0) \rangle,
\label{eq:dissipation-theorem-2}
\end{equation}
in deriving which we have used (\ref{eq:unrolling}).

The system is said to be in a nonequilibrium steady state
if the ensemble averages of all phase variables become 
time-independent. 
Let us notice that the long-time limit of (\ref{eq:dissipation-theorem-2})
becomes constant for systems	that exhibit mixing~\cite{comment-mixing}, 
which is to be assumed in the following. 
This feature can be demonstrated as follows.
By taking the time derivative of (\ref{eq:dissipation-theorem-2}), one finds
\begin{equation}
\frac{d}{dt} \langle A(t) \rangle =
\langle A(t) \Omega(0) \rangle. 
\label{eq:dA-over-dt}
\end{equation}
For systems that exhibit mixing~\cite{comment-mixing}, 
all the long-time correlations between phase variables vanish, and we obtain
for $t \to \infty$
\begin{equation}
\frac{d}{dt} \langle A(t) \rangle \to
\langle A(t) \rangle
\langle \Omega(0) \rangle = 0,
\label{eq:dA-over-dt-2}
\end{equation}
where we have used the property (\ref{eq:Omega-0}). 
This indicates that the long-time steady state average
of an arbitrary phase variable becomes constant,
i.e.,
\begin{equation}
\lim_{t \to \infty} \langle A(t) \rangle =
\langle A \rangle_{\rm ss},
\end{equation}
where the steady-state average, denoted by
$\langle \cdots \rangle_{\rm ss}$ hereafter,
is obtained from
the $t \to \infty$ limit of (\ref{eq:dissipation-theorem-2}):
\begin{eqnarray}
\langle A \rangle_{\rm ss} =
\langle A(0) \rangle 
+ \int_{0}^{\infty} ds \,
\langle A(s) \Omega(0) \rangle. 
\label{eq:ss-average}
\end{eqnarray}
This is called the generalized Green-Kubo relation
relating the steady-state average to the integral of the
transient time-correlation function~\cite{Evans08,Chong09b},
and applies to systems arbitrarily far from equilibrium. 
It reduces to the conventional Green-Kubo relation if the 
shearing force is weak and the viscous dissipative force is neglected.

\section{Cumulant-expansion representation of the distribution function}
\label{sec:cumulant-expansion-representation}

In this section, we derive a representation of the phase-space distribution
function which resembles the one developed in Refs.~\citen{Komatsu08} and \citen{Komatsu09},
but without assuming the microscopic time reversality.
Our derivation is certainly related, but not identical, to the one adopted in
Refs.~\citen{Komatsu08} and \citen{Komatsu09}. 

\subsection{Conditioned ``averages" and cumulant-expansion representation}

Let us first notice that, using the relation (\ref{eq:unrolling}) with
$\bv{\Gamma}(t) = e^{i {\cal L} t} \bv{\Gamma}$, 
the distribution function $f(\bv{\gamma},t)$ at a phase-space point $\bv{\gamma}$
can be expressed as
\begin{equation}
f({\bv \gamma},t) =
\int d\bv{\Gamma} \, f(\bv{\Gamma},t) \, \delta(\bv{\Gamma} - {\bv \gamma}) =
\int d\bv{\Gamma} \, f(\bv{\Gamma},0) \, \delta(\bv{\Gamma}(t) - {\bv \gamma}).
\label{eq:representation-f}
\end{equation}
With this notion, we shall introduce the conditioned ``average" $\langle C \rangle_{{\bv \gamma}; t}$
in which the phase-space point at time $t$ is constrained to $\bv{\gamma}$:
\begin{equation}
\langle C \rangle_{{\bv \gamma}; t} \equiv
\frac{\int d\bv{\Gamma} \, C \, f(\bv{\Gamma},0) \, \delta(\bv{\Gamma}(t) - {\bv \gamma})}
{\int d\bv{\Gamma} \,  f(\bv{\Gamma},0) \, \delta(\bv{\Gamma}(t) - {\bv \gamma})} =
\frac{\int d\bv{\Gamma} \, C \, f(\bv{\Gamma},0) \, \delta(\bv{\Gamma}(t) - {\bv \gamma})}
{f({\bv \gamma},t)}.
\label{eq:conditioned-average-t}
\end{equation}
Notice that $\langle C \rangle_{{\bv \gamma}; t}$ so defined is not really an
averaged quantity (this is why we put quotation marks) since, our equations 
of motion being deterministic,
it is completely determined by the single path in the phase-space that 
ends up at the point $\bv{\gamma}$ at time $t$.
We shall also introduce the conditioned average $\langle C \rangle_{{\bv \gamma}; 0}$, 
in which the phase-space point at time $t=0$ is constrained to $\bv{\gamma}$,
by setting $t=0$ in (\ref{eq:conditioned-average-t}):
\begin{equation}
\langle C \rangle_{{\bv \gamma}; 0} \equiv
\frac{\int d\bv{\Gamma} \, C \, f(\bv{\Gamma},0) \, \delta(\bv{\Gamma} - {\bv \gamma})}
{\int d\bv{\Gamma} \,  f(\bv{\Gamma},0) \, \delta(\bv{\Gamma} - {\bv \gamma})} =
\frac{\int d\bv{\Gamma} \, C \, f(\bv{\Gamma},0) \, \delta(\bv{\Gamma} - {\bv \gamma})}
{f({\bv \gamma},0)}.
\label{eq:conditioned-average-0}
\end{equation}

By inserting $C = e^{\frac{1}{2} \int_{0}^{t} ds \, \Omega(-s)}$ in 
(\ref{eq:conditioned-average-0}), we have 
\begin{eqnarray}
f({\bv \gamma},0) \, 
\langle e^{\frac{1}{2} \int_{0}^{t} ds \, \Omega(-s)} \rangle_{{\bv \gamma}; 0} &=&
\int d\tilde{\bv{\Gamma}} \, f(\tilde{\bv{\Gamma}},0) \,
e^{\frac{1}{2} \int_{0}^{t} ds \, \Omega(\tilde{\bv{\Gamma}}(-s))} \,
\delta(\tilde{\bv{\Gamma}}-{\bv \gamma}) 
\nonumber \\
&=&
\int d\tilde{\bv{\Gamma}} \, f(\tilde{\bv{\Gamma}},0) \,
e^{\frac{1}{2} \int_{-t}^{0} ds \, \Omega(\tilde{\bv{\Gamma}}(s))} \,
\delta(\tilde{\bv{\Gamma}}-{\bv \gamma}).
\end{eqnarray}
By setting $\tilde{\bv{\Gamma}} = \bv{\Gamma}(t)$,
one obtains, since $\tilde{\bv{\Gamma}}(s) = \bv{\Gamma}(t+s)$
\begin{equation}
f({\bv \gamma},0) \, 
\langle e^{\frac{1}{2} \int_{0}^{t} ds \, \Omega(-s)} \rangle_{{\bv \gamma}; 0} =
\int d\bv{\Gamma}(t) \, f(\bv{\Gamma}(t),0) \,
e^{\frac{1}{2} \int_{0}^{t} ds \, \Omega(\bv{\Gamma}(s))} \,
\delta(\bv{\Gamma}(t)-{\bv \gamma}). 
\end{equation}
Using (\ref{eq:Omega-property-2}), this leads to
\begin{eqnarray}
f({\bv \gamma},0) \, 
\langle e^{\frac{1}{2} \int_{0}^{t} ds \, \Omega(-s)} \rangle_{{\bv \gamma}; 0} &=&
\int d\bv{\Gamma} \, f(\bv{\Gamma},0) \,
e^{-\frac{1}{2} \int_{0}^{t} ds \, \Omega(\bv{\Gamma}(s))} \,
\delta(\bv{\Gamma}(t)-{\bv \gamma}) 
\nonumber \\
&=&
f({\bv \gamma},t) \, 
\langle e^{- \frac{1}{2} \int_{0}^{t} ds \, \Omega(s)} \rangle_{{\bv \gamma}; t},
\end{eqnarray}
where the second equality follows from the definition (\ref{eq:conditioned-average-t}).
We therefore obtain the following symmetrical
representation of the distribution function:
\begin{equation}
\frac{f({\bv \gamma},t)}{f({\bv \gamma},0)} =
\frac{\langle e^{\frac{1}{2} \int_{0}^{t} ds \, \Omega(-s)} \rangle_{{\bv \gamma}; 0}}
{\langle e^{- \frac{1}{2} \int_{0}^{t} ds \, \Omega(s)} \rangle_{{\bv \gamma}; t}} \, .
\label{eq:KN-like-expression-1}
\end{equation}

In the following, we shall manipulate the right-hand side of (\ref{eq:KN-like-expression-1})
using the cumulant expansion
\begin{equation}
\log \langle e^{Y} \rangle_{\bv{\gamma};t} = 
\sum_{k=1}^{\infty} \frac{1}{k!} \langle Y^{k} \rangle_{\bv{\gamma};t}^{\rm c}.
\label{eq:cumulant-expansion}
\end{equation}
Here $\langle Y^{k} \rangle_{\bv{\gamma};t}^{\rm c}$ denotes the $k$th-order cumulant of $Y$ defined by
\begin{equation}
\langle Y^{k} \rangle_{\bv{\gamma};t}^{\rm c} \equiv
\frac{\partial^{k}}{\partial u^{k}} \log \langle \exp[uY] \rangle_{\bv{\gamma};t} \Bigm|_{u=0}.
\label{eq:kth-cumulant-def}
\end{equation}
For example, $\langle Y \rangle_{\bv{\gamma};t}^{\rm c} = \langle Y \rangle_{\bv{\gamma};t}$ and
$\langle Y^{2} \rangle_{\bv{\gamma};t}^{\rm c} = \langle Y^{2} \rangle_{\bv{\gamma};t} - \langle Y \rangle_{\bv{\gamma};t}^{2}$. 
It is useful to note that for any constant $y_{0}$ there holds
\begin{equation}
\langle (Y - y_{0})^{k} \rangle_{\bv{\gamma};t}^{\rm c} = \langle Y^{k} \rangle_{\bv{\gamma};t}^{\rm c}
\,\,\, \mbox{for} \,\,\, k \ge 2,
\label{eq:cumulant-property}
\end{equation}
which can be derived on the basis of (\ref{eq:kth-cumulant-def}). 

By applying the cumulant expansion to (\ref{eq:KN-like-expression-1}), we obtain
\begin{equation}
\log \frac{f({\bv \gamma},t)}{f({\bv \gamma},0)} =
\frac{1}{2} 
\left\{
  \langle \Theta_{-} \rangle_{{\bv \gamma}; 0} +
  \langle \Theta_{+} \rangle_{{\bv \gamma}; t}
\right\} +
\sum_{k=2}^{\infty} \frac{1}{2^{k} k!}
\left\{
  \langle \Theta_{-}^{k} \rangle_{{\bv \gamma}; 0}^{\rm c} - (-1)^{k}
  \langle \Theta_{+}^{k} \rangle_{{\bv \gamma}; t}^{\rm c} 
\right\},
\label{eq:KN-like-expression-2}
\end{equation}
where we have introduced
\begin{equation}
\Theta_{-} \equiv \int_{0}^{t} ds \, \Omega(-s), \quad
\Theta_{+} \equiv \int_{0}^{t} ds \, \Omega(s).
\label{eq:Theta-def}
\end{equation}
Notice that the $t$ dependence is dropped from $\Theta_{\pm}$ for notational simplicity.

We would like to express (\ref{eq:KN-like-expression-2})
in terms of excess quantities.
For this purpose, one needs to introduce averages, so
let us define (only formally for the moment)
\begin{equation}
\bar{\Omega}_{-} \equiv
\lim_{t \to \infty} \frac{1}{t} \int_{0}^{t} ds \,
\langle \Omega(-s) \rangle_{{\bv \gamma};0}, \quad
\bar{\Omega}_{+} \equiv
\lim_{t \to \infty} \frac{1}{t} \int_{0}^{t} ds \,
\langle \Omega(s) \rangle_{{\bv \gamma};t}
\label{eq:bar-Omega-def}.
\end{equation}
Under what circumstances these limits exist (not in the mathematical sense but in the
physical sense) will be discussed below.
We will also argue below that these averages 
are independent of the choice of $\bv{\gamma}$, and this is why
we dropped the possible dependence on $\bv{\gamma}$ from 
the notation $\bar{\Omega}_{\pm}$.
Keeping these facets in mind, we shall introduce 
the excess quantities 
\begin{equation}
\Theta_{-}^{\rm ex} \equiv \int_{0}^{t} ds \, [ \, \Omega(-s) - \bar{\Omega}_{-} \, ], \quad
\Theta_{+}^{\rm ex} \equiv \int_{0}^{t} ds \,[ \,  \Omega(s) - \bar{\Omega}_{+} \, ].
\label{eq:excess-Theta-def}
\end{equation}
In view of the notice below (\ref{eq:Omega-property-2}),
$\Theta_{\pm}^{\rm ex}$ divided by $\beta$ can be identified as
the excess free energy production.
With the application of (\ref{eq:cumulant-property}), there hold
\begin{equation}
\langle \Theta_{-}^{k} \rangle_{{\bv \gamma};0}^{c} =
\langle (\Theta_{-}^{\rm ex})^{k} \rangle_{{\bv \gamma};0}^{c}, \quad
\langle \Theta_{+}^{k} \rangle_{{\bv \gamma};t}^{c} =
\langle (\Theta_{+}^{\rm ex})^{k} \rangle_{{\bv \gamma};t}^{c}
\,\,\, \mbox{for} \,\,\, k \ge 2
\label{eq:cumulant-property-2}
\end{equation}
for those cumulants in the second term on the right-hand side of (\ref{eq:KN-like-expression-2}).
For the first-order cumulants appearing there, we 
simply exploit the relation $\Theta_{\pm} = \Theta_{\pm}^{\rm ex} + t \bar{\Omega}_{\pm}$.
In this way, we obtain from (\ref{eq:KN-like-expression-2})
\begin{eqnarray}
\log \frac{f({\bv \gamma},t)}{f({\bv \gamma},0)} &=&
\frac{1}{2} 
\left\{
  \langle \Theta_{-}^{\rm ex} \rangle_{{\bv \gamma}; 0} +
  \langle \Theta_{+}^{\rm ex} \rangle_{{\bv \gamma}; t}
\right\} + 
\frac{t}{2} \, ( \bar{\Omega}_{-} + \bar{\Omega}_{+}) 
\nonumber \\
& & \qquad
+ \,
\sum_{k=2}^{\infty} \frac{1}{2^{k} k!}
\left\{
  \langle (\Theta_{-}^{\rm ex})^{k} \rangle_{{\bv \gamma}; 0}^{\rm c} - (-1)^{k}
  \langle (\Theta_{+}^{\rm ex})^{k} \rangle_{{\bv \gamma}; t}^{\rm c} 
\right\},
\label{eq:KN-like-expression-3}
\end{eqnarray}
which is the starting point of the following formulation.

\subsection{Expansion in the small ``degree of nonequilibrium"}
\label{subsec:expansion}

The representation (\ref{eq:KN-like-expression-3}) is formal, and will be useful in practice
only if it is dominated by a first few terms in the expansion.
We shall therefore explore in the following the form of the distribution function
by restricting ourselves to the 
small ``degree of nonequilibrium" to be characterized by a small
parameter $\epsilon$.
We then examine the order in $\epsilon$ of each term on the right-hand side of (\ref{eq:KN-like-expression-3}). 
The usefulness of such an expansion has been demonstrated in Ref.~\citen{Komatsu08b},
where the distribution function valid to the second order in $\epsilon$
has been exploited to derive
thermodynamic relations for nonequilibrium steady states.

As we have noticed above, not only the shearing force, but also the
viscous dissipative force should be considered 
as the driving force towards nonequilibrium states.
The magnitude of the shearing force is determined by the shear rate $\dot{\gamma}$,
whereas from (\ref{eq:F-vis-def}) that of the viscous dissipative force is characterized by 
$\zeta/m$ assuming a constant viscous function, $\zeta(x) = \zeta$.
The small degree of nonequilibrium shall therefore be characterized by 
small $\dot{\gamma}$ and $\zeta/m$,
and the parameter $\epsilon$ shall be chosen as the larger of $\dot{\gamma}$ and $\zeta/m$
so that the correction $i \Delta {\cal L}$ to the unperturbed Liouvillean $i{\cal L}_{0}$ in 
(\ref{eq:explicit-iLp}) reads $i \Delta {\cal L} = O(\epsilon)$, i.e.,
\begin{equation}
i {\cal L} = i {\cal L}_{0} + i \Delta {\cal L}
\,\,\, \mbox{with} \,\,\, i \Delta {\cal L} = O(\epsilon).
\label{eq:iLp-expansion}
\end{equation}
Notice that our characterization of the small degree of nonequilibrium makes sense
since small $\zeta/m$ (i.e., quasielastic limit) is involved there.
At the same time, we have to assume small $\dot{\gamma}$ since our primary interest here
is in the transient dynamics occurring before the steady state is reached.
These points are discussed in Appendix~\ref{appendix:Balance}.
We also note that $\epsilon$ so chosen has the dimension of inverse of time.
This is because, as we will see below, $\epsilon$ appears mostly in the form
of the product $\epsilon \tau$ with the relaxation time $\tau$.
It is therefore more preferable to have $\epsilon \tau$ dimensionless.

Similarly to (\ref{eq:iLp-expansion}), we have from (\ref{eq:SLLOD-Lambda}) and (\ref{eq:explicit-iLf})
\begin{equation}
i {\cal L}^{\dagger} = i {\cal L}_{0}^{\dagger} + i \Delta {\cal L}^{\dagger}
\,\,\, \mbox{with} \,\,\, 
i \Delta {\cal L}^{\dagger} = O(\epsilon).
\label{eq:iLf-expansion}
\end{equation}
Correspondingly, $p$- and $f$-propagators can be expanded as
\begin{equation}
e^{i {\cal L} t} = e^{i {\cal L}_{0} t} + O(\epsilon t), \quad
e^{-i {\cal L}^{\dagger} t} = e^{-i {\cal L}_{0}^{\dagger} t} + O(\epsilon t).
\label{eq:propagator-expansion}
\end{equation}
It is also clear from (\ref{eq:SLLOD-Lambda}) and (\ref{eq:Omega-def}) that
\begin{equation}
\Omega = O(\epsilon),
\label{eq:Omega-epsilon}
\end{equation}
which plays an important role in the following.
For later convenience, we shall introduce the notation
$\bv{\Gamma}_{0}(t) \equiv e^{i {\cal L}_{0} t} \bv{\Gamma}$
evolving under the ``unperturbed" $p$-propagator $e^{i {\cal L}_{0} t}$.
Because of (\ref{eq:propagator-expansion}), there hold
\begin{equation}
\bv{\Gamma}(t) = \bv{\Gamma}_{0}(t) + O(\epsilon t), \quad
A(\bv{\Gamma}(t)) = A(\bv{\Gamma}_{0}(t)) + O(A) \cdot O(\epsilon t).
\label{eq:Gamma-A-epsilon}
\end{equation}

Since the initial distribution function is chosen as the solution of
$i {\cal L}_{0}^{\dagger}(\bv{\Gamma}) f(\bv{\Gamma},0) = 0$, it follows
from the relation (\ref{eq:iL-dagger-fA}) specialized to 
the Liouvilleans $i{\cal L}_{0}^{\dagger}$ and $i {\cal L}_{0}$ that
\begin{equation}
i {\cal L}^{\dagger}_{0}(\bv{\Gamma}) [ f(\bv{\Gamma},0) A(\bv{\Gamma}) ] =
f(\bv{\Gamma},0) [ i {\cal L}_{0}(\bv{\Gamma}) A(\bv{\Gamma})].
\label{eq:iL0-fA-1}
\end{equation}
By repeated applications of this result, one obtains
\begin{equation}
e^{i {\cal L}_{0}^{\dagger}(\bv{\Gamma}) t} \, [ f(\bv{\Gamma},0) A(\bv{\Gamma}) ] =
f(\bv{\Gamma},0) \, [ e^{i {\cal L}_{0}(\bv{\Gamma}) t} A(\bv{\Gamma}) ],
\label{eq:iL0-fA-2}
\end{equation}
which is to be used in the following.

\subsection{Order estimate}
\label{subsec:order-estimate}

To estimate the order in $\epsilon$ of each term on the right-hand side of (\ref{eq:KN-like-expression-3}),
one needs to estimate the order of $\Theta_{\pm}^{\rm ex}$.
For this purpose, we need to know more about the averages $\bar{\Omega}_{\pm}$ 
defined in (\ref{eq:bar-Omega-def}).
Let us start from a related quantity
\begin{equation}
\lim_{t \to \infty} \frac{1}{t} \int_{0}^{t} ds \,
\langle \Omega(s) \rangle_{{\bv \gamma};0} =
\lim_{t \to \infty} \frac{1}{t} \int_{0}^{t} ds \,
\Omega({\bv \gamma}(s)).
\label{eq:bar-Omega-dum-01}
\end{equation}
The system is expected to settle to a unique nonequilibrium steady state,
irrespective of the initial phase-space point $\bv{\gamma}$,
for times longer than the relaxation time which we denote as $\tau$. 
Here, $\tau$ shall be chosen so that the system reaches the steady
state for $t \gtrsim \tau$. 
Then, by writing
\begin{equation}
\int_{0}^{t} ds \, \Omega({\bv \gamma}(s)) =
\int_{0}^{\tau} ds \, \Omega({\bv \gamma}(s)) +
\int_{\tau}^{t} ds \, \Omega({\bv \gamma}(s)),
\end{equation}
the second term on the right-hand side can be estimated as
$(t-\tau) \langle \Omega \rangle_{\rm ss}$ 
with the steady-state average
$\langle \Omega \rangle_{\rm ss}$. 
As a result, there holds irrespective of the choice of $\bv{\gamma}$
\begin{equation}
\lim_{t \to \infty} \frac{1}{t} \int_{0}^{t} ds \,
\langle \Omega(s) \rangle_{{\bv \gamma};0} =
\langle \Omega \rangle_{\rm ss}.
\label{eq:bar-Omega-plus-0}
\end{equation}

Now, we apply this argument to 
$\bar{\Omega}_{+} = \lim_{t \to \infty} (1/t) \int_{0}^{t} ds \,
\langle \Omega(s) \rangle_{{\bv \gamma};t}$.
The difference from (\ref{eq:bar-Omega-dum-01}) is that here what is fixed is the
phase-space point $\bv{\gamma}$ at time $t$.
To take this into account, we introduce the time $\tau^{\prime}$ required
to reach the phase-space point $\bv{\gamma}$ during the steady-state fluctuations.
Then, the effect of fixing the end point $\bv{\gamma}$ at time $t$
will show up only in the time regime $t - \tau^{\prime} \lesssim s \le t$.
Therefore, by writing
\begin{equation}
\int_{0}^{t} ds \, \langle \Omega(s) \rangle_{{\bv \gamma},t} =
\int_{0}^{\tau} ds \, \langle \Omega(s) \rangle_{{\bv \gamma},t} +
\int_{\tau}^{t-\tau^{\prime}} ds \, \langle \Omega(s) \rangle_{{\bv \gamma},t} +
\int_{t-\tau^{\prime}}^{t} ds \, \langle \Omega(s) \rangle_{{\bv \gamma},t},
\end{equation}
and noticing that the second term on the right-hand side
can be estimated as $(t-\tau^{\prime}-\tau) \langle \Omega \rangle_{\rm ss}$,
there holds irrespective of the choice of ${\bv \gamma}$
\begin{equation}
\bar{\Omega}_{+} =
\lim_{t \to \infty} \frac{1}{t} \int_{0}^{t} ds \,
\langle \Omega(s) \rangle_{{\bv \gamma};t} = 
\langle \Omega \rangle_{\rm ss} =
\lim_{t \to \infty} \langle \Omega(t) \rangle.
\label{eq:bar-Omega-plus}
\end{equation}
From this argument, it is also clear that the integrand of
$\Theta_{+}^{\rm ex} = \int_{0}^{t} ds \, [ \Omega(s) - \bar{\Omega}_{+}]$
is non-negligible only for $0 \le s \lesssim \tau$ and $t-\tau^{\prime} \lesssim s \le t$.
As will be discussed in the next subsection, however, we will eventually set the upper limit
$t$ of the integral in $\Theta_{+}^{\rm ex}$ to $\tau$.
In this case, since $\Omega = O(\epsilon)$, there holds
\begin{equation}
\Theta_{+}^{\rm ex} = O(\epsilon \tau).
\label{eq:estimate-excess-Theta-plus}
\end{equation}

One can apply a similar argument to $\bar{\Omega}_{-}$ and $\Theta_{-}^{\rm ex}$, 
assuming that a steady-state is reached in the negative time direction
(see below concerning this point), to obtain
\begin{equation}
\bar{\Omega}_{-} =
\lim_{t \to \infty} \frac{1}{t} \int_{0}^{t} ds \,
\langle \Omega(-s) \rangle_{{\bv \gamma};0} =
\lim_{t \to \infty} \langle \Omega(-t) \rangle,
\label{eq:bar-Omega-minus}
\end{equation}
irrespective of the choice of $\bv{\gamma}$, and
\begin{equation}
\Theta_{-}^{\rm ex} =  O(\epsilon \tau).
\label{eq:estimate-excess-Theta-minus}
\end{equation}

From (\ref{eq:estimate-excess-Theta-plus}) and (\ref{eq:estimate-excess-Theta-minus}),
one understands that the cumulants of $k \ge 3$ appearing in the right-hand side of 
(\ref{eq:KN-like-expression-3}) are of order $O(\epsilon^{3} \tau^{3})$.
In addition, one finds that the second-order cumulant term
$\langle (\Theta_{-}^{\rm ex})^{2} \rangle_{{\bv \gamma}; 0}^{\rm c} - 
\langle (\Theta_{+}^{\rm ex})^{2} \rangle_{{\bv \gamma}; t}^{\rm c}$
is also $O(\epsilon^{3} \tau^{3})$,
and this is demonstrated in Appendix~\ref{appendix:B}.
We therefore obtain from (\ref{eq:KN-like-expression-3})
\begin{equation}
\log \frac{f({\bv \gamma},t)}{f({\bv \gamma},0)} =
\frac{1}{2} 
\left\{
  \langle \Theta_{-}^{\rm ex} \rangle_{{\bv \gamma}; 0} +
  \langle \Theta_{+}^{\rm ex} \rangle_{{\bv \gamma}; t}
\right\} + 
\frac{t}{2} \, ( \bar{\Omega}_{-} + \bar{\Omega}_{+}) + O(\epsilon^{3} \tau^{3}).
\label{eq:KN-like-expression-4}
\end{equation}
Now, what is left is the order estimate of the second term, 
which we write in view of (\ref{eq:bar-Omega-plus}) and (\ref{eq:bar-Omega-minus}) as
\begin{equation}
\bar{\Omega}_{-} + \bar{\Omega}_{+} = \lim_{t \to \infty} 
\{ \langle \Omega(-t) \rangle + \langle \Omega(t) \rangle \}.
\label{eq:bar-Omega-sum}
\end{equation}
For systems possessing the time-reversal symmetry, one can show that
$\bar{\Omega}_{-} + \bar{\Omega}_{+} = 0$.
However, this is not the case for systems in which such a symmetry is broken.

\subsection{On the relation between $\bar{\Omega}_{+}$ and $\bar{\Omega}_{-}$}
\label{subsec:Omega-pm-time-irreversal}

Here, we shall deal with $\bar{\Omega}_{\pm}$ for systems 
in which microscopic time-reversal symmetry is broken.
The average $\bar{\Omega}_{+} = \lim_{t \to \infty} \langle \Omega(t) \rangle$ in the positive
time direction is well defined since the system does reach the steady state for $t \gtrsim \tau$.
However, in general, the average
$\bar{\Omega}_{-} = \lim_{t \to \infty} \langle \Omega(-t) \rangle$
defined in the negative time direction does exist for 
the time-irreversible system. 
This is clear for the sheared granular system under study:
in the negative time direction, 
particles ``attain" energy upon collisions, and the system continuously heats up so that
the steady state is never reached. 
However, by considering the small degree $\epsilon$ of nonequilibrium, 
there might be a possibility that such a heat up of the system up to the time scale $-\tau$
in the negative direction is still negligible.
We shall therefore explore in the following 
the relation between $\bar{\Omega}_{+}$ and $\bar{\Omega}_{-}$ 
that holds for small $\epsilon$ and up to the time scale $\pm \tau$.
This means that, instead of (\ref{eq:bar-Omega-sum}), 
we will consider
\begin{equation}
\bar{\Omega}_{-} + \bar{\Omega}_{+} \approx
\{ \langle \Omega(-t) \rangle + \langle \Omega(t) \rangle \} \Bigm|_{t = \tau}
\label{eq:bar-Omega-sum-at-tau}
\end{equation}
evaluated at time $\tau$ just after the steady state is reached.
This is the reason why we eventually set the upper limit $t$ of the integral 
in $\Theta_{\pm}^{\rm ex}$ to $\tau$, which is assumed above in 
connection with (\ref{eq:estimate-excess-Theta-plus}).

We start from the generalized Green-Kubo relation (\ref{eq:dissipation-theorem-2}) for 
$\langle \Omega(-t) \rangle$ and $\langle \Omega(t) \rangle$.
Since $\langle \Omega(0) \rangle = 0$, we have
\begin{eqnarray}
& &
\langle \Omega(-t) \rangle =
\int_{0}^{-t} ds \, \langle \Omega(s) \Omega(0) \rangle =
- \int_{0}^{t} ds \, \langle \Omega(-s) \Omega(0) \rangle,
\label{eq:gGK-Omega-minus-def}
\\
& &
\langle \Omega(t) \rangle =
\int_{0}^{t} ds \, \langle \Omega(s) \Omega(0) \rangle.
\label{eq:gGK-Omega-plus-def}
\end{eqnarray}
Using the decomposition (\ref{eq:propagator-expansion}) for $e^{i{\cal L}t}$ and noticing 
$\Omega = O(\epsilon)$,
one finds for $\langle \Omega(t) \rangle$:
\begin{eqnarray}
\langle \Omega(t) \rangle &=&
\int_{0}^{t} ds \int d\bv{\Gamma} \,
f(\bv{\Gamma},0) 
\left[ e^{i {\cal L} s} \Omega(\bv{\Gamma}) \right]
\Omega(\bv{\Gamma})
\nonumber \\
&=&
\int_{0}^{t} ds \int d\bv{\Gamma} \,
f(\bv{\Gamma},0) 
\left[ \left\{e^{i {\cal L}_{0} s} + O(\epsilon s) \right\} \Omega(\bv{\Gamma}) \right]
\Omega(\bv{\Gamma})
\nonumber \\
&=&
\int_{0}^{t} ds \int d\bv{\Gamma} \,
f(\bv{\Gamma},0) 
\left[ e^{i {\cal L}_{0} s} \Omega(\bv{\Gamma}) \right]
\Omega(\bv{\Gamma}) + O(\epsilon^{3} t^{2}).
\label{eq:gGK-Omega-plus-1}
\end{eqnarray}
Similarly, one obtains for $\langle \Omega(-t) \rangle$:
\begin{eqnarray}
\langle \Omega(-t) \rangle &=&
- \int_{0}^{t} ds \int d\bv{\Gamma} \,
f(\bv{\Gamma},0) 
\left[ e^{- i {\cal L} s} \Omega(\bv{\Gamma}) \right] 
\Omega(\bv{\Gamma})
\nonumber \\
&=&
- \int_{0}^{t} ds \int d\bv{\Gamma} \,
f(\bv{\Gamma},0) 
\left[ e^{- i {\cal L}_{0} s} \, \Omega(\bv{\Gamma}) \right] 
\Omega(\bv{\Gamma}) + O(\epsilon^{3} t^{2}).
\label{eq:gGK-Omega-minus-1}
\end{eqnarray}
Applying first (\ref{eq:unrolling}) and then (\ref{eq:iL0-fA-2}) to this expression, one finds
\begin{eqnarray}
\langle \Omega(-t) \rangle &=&
- \int_{0}^{t} ds \int d\bv{\Gamma} \,
f(\bv{\Gamma},0) \, 
\Omega(\bv{\Gamma})
\left[ e^{i {\cal L}_{0} s} \Omega(\bv{\Gamma}) \right] + O(\epsilon^{3} t^{2})
\nonumber \\
&=&
- \langle \Omega(t) \rangle + O(\epsilon^{3} t^{2}).
\label{eq:gGK-Omega-minus-2}
\end{eqnarray}
Thus, from the identification (\ref{eq:bar-Omega-sum-at-tau}), we have for small $\epsilon$
\begin{equation}
t \, (\bar{\Omega}_{-} + \bar{\Omega}_{+}) \Bigm|_{t = \tau} = O(\epsilon^{3} \tau^{3}),
\label{eq:bar-Omega-sum-irreversible}
\end{equation}
for the time-irreversible system at time $\tau$
just after the steady state is reached.

\section{Summary}
\label{sec:summary}

Collecting our results so far, we have for the steady-state
distribution function $f_{\rm ss}(\bv{\Gamma})$ for the small degree $\epsilon$ of nonequilibrium
\begin{equation}
f_{\rm ss}(\bv{\Gamma}) = f(\bv{\Gamma},0)
\exp \Bigl[ \,
  \frac{1}{2}\left\{
  \langle \Theta_{-}^{\rm ex} \rangle_{{\bv \Gamma}; 0} +
  \langle \Theta_{+}^{\rm ex} \rangle_{{\bv \Gamma}; t}
  \right\} \,
\Bigr] + O(\epsilon^{3} \tau^{3}).
\label{eq:KN-like-expression}
\end{equation}
As argued above, 
$f_{\rm ss}(\bv{\Gamma})$ for the system in which time-reversal symmetry is broken
should be considered as the one just after the system entered the steady state at time $\tau$,
$f_{\rm ss}(\bv{\Gamma}) = f(\bv{\Gamma},t) |_{t = \tau}$.
Up to the presence of $\tau$ in the correction terms,
one understands that the representation (\ref{eq:KN-like-expression}) for the
steady-state distribution function is essentially the same as the one
derived in Refs.~\citen{Komatsu08} and \citen{Komatsu09} in that
it is expressed in terms of the excess free energy productions 
and that it is valid up to the second order in $\epsilon$, i.e., it holds
beyond the linear-response regime.

However, there is a subtle point connected with the presence of $\tau$ 
in the correction terms.
In fact, there holds $\epsilon \tau = O(1)$ on general grounds.
This can be understood from the generalized Green-Kubo relation 
$\langle \Omega(t) \rangle = \int_{0}^{t} ds \,
\langle \Omega(s) \Omega(0) \rangle$
by noticing that the left-hand side is $O(\epsilon)$ whereas
the right-hand side is $O(\epsilon^{2} \tau)$ since $\langle \Omega(s) \Omega(0) \rangle$
decays to zero on the time scale of $\tau$.
Thus, all the correction terms in the representation (\ref{eq:KN-like-expression})
are of order $O(1)$, and hence, it is not a converging representation.
In this sense, the nonequilibrium steady-state distribution function might not exist.
This point will be discussed in more detail in our subsequent publication.

\section*{Acknowledgements}

We thank S.~Sasa and H.~Tasaki for discussions. 
This work was supported by the Grant-in-Aid for scientific
research from the Ministry of Education, Culture, Sports,
Science and Technology (MEXT) of Japan 
(Nos.~20740245, 21015016, 21540384, and 21540388),
by the Global COE Program
``The Next Generation of Physics, Spun from Universality
and Emergence'' from MEXT of Japan,
and in part by the Yukawa International Program for 
Quark-Hadron Sciences (YIPQS).

\appendix

\section{Order estimate of the second-order cumulant term}
\label{appendix:B}

In this appendex, we show that
\begin{equation}
\langle (\Theta_{-}^{\rm ex})^{2} \rangle_{{\bv \gamma};0}^{\rm c} - 
\langle (\Theta_{+}^{\rm ex})^{2} \rangle_{{\bv \gamma};t}^{\rm c} = O(\epsilon^{3} \tau^{3}).
\label{eq:dum-01}
\end{equation}
To this end, we notice that with the remark given above (\ref{eq:estimate-excess-Theta-plus})
there holds 
$\Theta_{\pm}^{\rm ex} = O(\epsilon \tau)$ in leading order for small $\epsilon$.
Therefore, what (\ref{eq:dum-01}) claims is that such a leading-order contribution 
in the left-hand side cancels out.

Using the notation in (\ref{eq:Gamma-A-epsilon}), 
the leading-order term is given by
\begin{equation}
\Theta_{\pm \, 0}^{\rm ex} \equiv
\int_{0}^{t} ds \, [ \Omega(\bv{\Gamma}_{0}(\pm s)) - \bar{\Omega}_{\pm \, 0}].  
\label{eq:dum-05}
\end{equation}
Here, $\bar{\Omega}_{\pm \, 0}$ is defined as in (\ref{eq:bar-Omega-def})
but with $e^{\pm i {\cal L} s}$ replaced by the
unperturbed $e^{\pm i {\cal L}_{0} s}$. 
Then, to show the validity of (\ref{eq:dum-01}), it suffices to demonstrate
\begin{equation}
\langle (\Theta_{- \, 0}^{\rm ex})^{2} \rangle_{{\bv \gamma};0}^{\rm c} - 
\langle (\Theta_{+ \, 0}^{\rm ex})^{2} \rangle_{{\bv \gamma};\tau}^{\rm c} = 0.
\label{eq:dum-06}
\end{equation}
Using the property (\ref{eq:cumulant-property}) 
and expanding cumulants in terms of moments, this means that
what we have to show are the equalities
\begin{eqnarray}
& &
\int_{0}^{t} ds \, 
\langle \Omega(-s) \rangle_{{\bv \gamma};0} =
\int_{0}^{t} ds \, 
\langle \Omega(s) \rangle_{{\bv \gamma};t},
\label{eq:dum-21}
\\
& &
\int_{0}^{t} ds \int_{0}^{t} ds' \, 
\langle \Omega(-s) \Omega(-s') \rangle_{{\bv \gamma};0} =
\int_{0}^{t} ds \int_{0}^{t} ds' \, 
\langle \Omega(s) \Omega(s') \rangle_{{\bv \gamma};t},
\label{eq:dum-22}
\end{eqnarray}
in which $p$- and $f$-Liouvilleans are given by $i{\cal L}_{0}$ and $i{\cal L}_{0}^{\dagger}$, respectively.

We start from deriving (\ref{eq:dum-21}).
From the definition (\ref{eq:conditioned-average-0}) of the conditioned average,
the left-hand side of (\ref{eq:dum-21}) is given by
\begin{equation}
\langle \Omega(-s) \rangle_{{\bv \gamma};0} =
\frac{1}{f({\bv \gamma},0)}
\int_{0}^{t} ds \int d\bv{\Gamma} \,
\left[ e^{-i{\cal L}_{0}(\bv{\Gamma})s} \Omega(\bv{\Gamma}) \right] 
f(\bv{\Gamma},0) \delta(\bv{\Gamma}-{\bv \gamma}).
\end{equation}
Applying first (\ref{eq:unrolling}) and then (\ref{eq:iL0-fA-2})
to this expression, we obtain
\begin{equation}
\int_{0}^{t} ds \, 
\langle \Omega(-s) \rangle_{{\bv \gamma};0} =
\frac{1}{f({\bv \gamma},0)}
\int_{0}^{t} ds \int d\bv{\Gamma} \,
\Omega(\bv{\Gamma}) \, f(\bv{\Gamma},0) \, \delta(\bv{\Gamma}_{0}(s)-{\bv \gamma}).
\label{eq:dum-23}
\end{equation}
On the other hand, from the definition (\ref{eq:conditioned-average-t}) of the conditioned average,
the right-hand side of (\ref{eq:dum-21}) is given by
\begin{equation}
\int_{0}^{t} ds \, 
\langle \Omega(s) \rangle_{{\bv \gamma};t} =
\frac{1}{f({\bv \gamma},0)}
\int_{0}^{t} ds \int d\bv{\Gamma} \,
\left[ e^{i {\cal L}_{0}(\bv{\Gamma}) s} \Omega(\bv{\Gamma}) \right] f(\bv{\Gamma},0) \delta(\bv{\Gamma}_{0}(t)-{\bv \gamma}).
\end{equation}
Here we have noticed that $f(\bv{\gamma},t) = f(\bv{\gamma},0)$ when the $f$-Liouvillean is given
by $i{\cal L}_{0}^{\dagger}$.
Again, by applying first (\ref{eq:unrolling}) and then (\ref{eq:iL0-fA-2})
to this expression, one obtains
\begin{eqnarray}
\int_{0}^{t} ds \, 
\langle \Omega(s) \rangle_{{\bv \gamma};t} &=&
\frac{1}{f({\bv \gamma},0)}
\int_{0}^{t} ds \int d\bv{\Gamma} \,
\Omega(\bv{\Gamma}) f(\bv{\Gamma},0) \delta(\bv{\Gamma}_{0}(t-s)-{\bv \gamma}) 
\nonumber \\
&=& 
\frac{1}{f({\bv \gamma},0)}
\int_{0}^{t} ds \int d\bv{\Gamma} \,
\Omega(\bv{\Gamma}) \, f(\bv{\Gamma},0) \, \delta(\bv{\Gamma}_{0}(s)-{\bv \gamma}),
\label{eq:dum-24}
\end{eqnarray}
and thus the desired equality (\ref{eq:dum-21}) is derived. 

One can derive the equality (\ref{eq:dum-22}) in a similar manner,
so we only show the course of the derivation.
First, the left-hand side of (\ref{eq:dum-22}) can be manipulated as
\begin{eqnarray}
& &
\int_{0}^{t} ds \int_{0}^{t} ds' \, 
\langle \Omega(-s) \Omega(-s') \rangle_{{\bv \gamma};0} 
\nonumber \\
& & 
=
\frac{1}{f({\bv \gamma},0)}
\int_{0}^{t} ds \int_{0}^{t} ds' \int d\bv{\Gamma} \,
\left[ e^{-i{\cal L}_{0}(\bv{\Gamma}) s} \Omega(\bv{\Gamma}) \right]
\left[ e^{-i{\cal L}_{0}(\bv{\Gamma}) s'} \Omega(\bv{\Gamma}) \right] f(\bv{\Gamma},0)
\delta(\bv{\Gamma} - {\bv \gamma})
\nonumber \\
& & 
=
\frac{1}{f({\bv \gamma},0)}
\int_{0}^{t} ds \int_{0}^{t} ds' \int d\bv{\Gamma} \,
\Omega(\bv{\Gamma}) \Omega(\bv{\Gamma}_{0}(s-s')) f(\bv{\Gamma},0) \,
\delta(\bv{\Gamma}_{0}(s) - {\bv \gamma}).
\label{eq:dum-25}
\end{eqnarray}
For the right-hand side of (\ref{eq:dum-22}), one can proceed as follows:
\begin{eqnarray}
& &
\int_{0}^{t} ds \int_{0}^{t} ds' \, 
\langle \Omega(s) \Omega(s') \rangle_{{\bv \gamma};t}
\nonumber \\
& & 
=
\frac{1}{f({\bv \gamma},0)}
\int_{0}^{t} ds \int_{0}^{t} ds' \int d\bv{\Gamma} \,
\left[ e^{i {\cal L}_{0}(\bv{\Gamma}) s} \Omega(\bv{\Gamma}) \right] 
\left[ e^{i {\cal L}_{0}(\bv{\Gamma}) s'} \Omega(\bv{\Gamma}) \right] f(\bv{\Gamma},0) 
\delta(\bv{\Gamma}_{0}(t) - {\bv \gamma})
\nonumber \\
& & 
=
\frac{1}{f({\bv \gamma},0)}
\int_{0}^{t} ds \int_{0}^{t} ds' \int d\bv{\Gamma} \,
\Omega(\bv{\Gamma}) \Omega(\bv{\Gamma}_{0}(s'-s)) f(\bv{\Gamma},0) 
\delta(\bv{\Gamma}_{0}(t-s) - {\bv \gamma})
\nonumber \\
& & 
=
\frac{1}{f({\bv \gamma},0)}
\int_{0}^{t} ds \int_{0}^{t} ds' \int d\bv{\Gamma} \,
\Omega(\bv{\Gamma}) \Omega(\bv{\Gamma}_{0}(s'-t+s)) f(\bv{\Gamma},0)
\delta(\bv{\Gamma}_{0}(s) - {\bv \gamma})
\nonumber \\
& & 
=
\frac{1}{f({\bv \gamma},0)}
\int_{0}^{t} ds \int_{0}^{t} ds' \int d\bv{\Gamma} \,
\Omega(\bv{\Gamma}) \Omega(\bv{\Gamma}_{0}(s-s')) f(\bv{\Gamma},0) 
\delta(\bv{\Gamma}_{0}(s) - {\bv \gamma}). 
\label{eq:dum-26}
\end{eqnarray}
In this way, the desired equality (\ref{eq:dum-22}) is derived.

\section{Balance equation for the kinetic temperature}
\label{appendix:Balance}

In this appendix, we shall derive a balance equation for the kinetic temperature
$T(t) \equiv 2 \langle K(t) \rangle / (DN)$ defined for a $D$-dimensional system in terms of the
average of the kinetic energy $K = \sum_{i} \bv{p}_{i}^{2}/(2m)$ at time $t$, and discuss
its implication. 
We first notice that the time derivative of $K$ is given by
\begin{equation}
\dot{K} =
\sum_{i} \frac{\bv{p}_{i}}{m} \cdot \dot{\bv{p}}_{i} =
\sum_{i} \frac{\bv{p}_{i}}{m} \cdot \Bigl( \bv{F}_{i}^{\rm (el)} + \bv{F}_{i}^{\rm (vis)} \Bigr) -
\bv{\kappa} : \sum_{i} \bv{p}_{i} \bv{p}_{i} / m,
\end{equation}
where we have used (\ref{eq:SLLOD-b}) and 
introduced the notation $\bv{A} : \bv{B} \equiv \sum_{\alpha, \beta} A_{\alpha \beta} B_{\beta \alpha}$.
Using $\bv{p}_{i}/m = \dot{\bv{r}}_{i} - \bv{\kappa} \cdot \bv{r}_{i}$
from (\ref{eq:SLLOD-a}), one obtains
\begin{eqnarray}
\dot{K} &=&
\sum_{i} \dot{\bv{r}}_{i} \cdot \Bigl( \bv{F}_{i}^{\rm (el)} + \bv{F}_{i}^{\rm (vis)} \Bigr) -
\bv{\kappa} : \sum_{i} \Bigl(  \bv{p}_{i} \bv{p}_{i} / m + \bv{r}_{i} \bv{F}_{i}^{\rm (el)} + \bv{r}_{i} \bv{F}_{i}^{\rm (vis)} \Bigr)
\nonumber \\
&=&
\sum_{i} \dot{\bv{r}}_{i} \cdot \bv{F}_{i}^{\rm (el)} +
\frac{1}{2} \sum_{i,j} \bv{g}_{ij} \cdot \bv{F}_{ij}^{\rm (vis)} -
\bv{\kappa} : \bv{\sigma}.
\label{eq:K-dot-dum-2}
\end{eqnarray}
In the second equality, we have used 
$\bv{g}_{ij} = \dot{\bv{r}}_{i} - \dot{\bv{r}}_{j}$, 
$\bv{F}_{i}^{\rm (vis)} = \sum_{j \ne i} \bv{F}_{ij}^{\rm (vis)}$, 
Newton's third law
$\bv{F}_{ji}^{\rm (vis)} = -\bv{F}_{ij}^{\rm (vis)}$, and
the definition (\ref{eq:Omega-def-b}) of the shear-stress tensor.
The first term on the right-hand side can be related to the time derivative $\dot{U}$
of the total potential energy since 
$\dot{U} = \sum_{i} \dot{\bv{r}}_{i} \cdot (\partial U/\partial \bv{r}_{i})$ and
$\bv{F}_{i}^{\rm (el)} = - \partial U / \partial \bv{r}_{i}$.
The second term can be expressed 
in terms of Rayleigh's dissipation function since 
${\cal R} = -(1/4) \sum_{i,j} \bv{g}_{ij} \cdot \bv{F}_{ij}^{\rm (vis)}$
(see (\ref{eq:F-vis-def}) and (\ref{eq:Omega-def-c})).
We therefore obtain from (\ref{eq:K-dot-dum-2}) 
\begin{equation}
\dot{K} = - \dot{U} - \dot{\gamma} \sigma_{xy} - 2 {\cal R},
\label{eq:K-dot-dum-3}
\end{equation}
where we have used the explicit form 
$\kappa_{\alpha \beta} = \dot{\gamma} \delta_{\alpha x} \delta_{\beta y}$
of the shear-rate tensor.

To simplify the following discussion, we assume a constant viscous function,
$\zeta(x) = \zeta$.
Then, $\zeta$ can be factored out from ${\cal R}$ (see (\ref{eq:Omega-def-c})), 
and we define ${\cal D}(t)$ via
$\zeta {\cal D}(t) \equiv \langle {\cal R}(t) \rangle / V$.
Let us also introduce the nonlinear viscosity $\eta(t)$ via
$\dot{\gamma} \eta(t) \equiv - \langle \sigma_{xy}(t) \rangle / V$.
One then obtains from (\ref{eq:K-dot-dum-3})
the following time-evolution equation, or the balance equation, for the kinetic temperature $T(t)$:
\begin{equation}
\dot{T}(t) = 
- \frac{2}{DN} \langle \dot{U}(t) \rangle +
\frac{2}{Dn} \dot{\gamma}^{2} \eta(t) - \frac{4}{Dn} \zeta {\cal D}(t).
\label{eq:K-dot-dum-4}
\end{equation}
Here $n \equiv N/V$ denotes the average number density.

As discussed in connection with (\ref{eq:dA-over-dt-2}),
the time derivative of the average of any phase-space variable vanishes in the
steady state.
Thus, there holds from (\ref{eq:K-dot-dum-4})
\begin{equation}
\dot{T}(t) \to 
\frac{2}{Dn} \dot{\gamma}^{2} \eta_{\rm ss} - \frac{4}{Dn} \zeta {\cal D}_{\rm ss} = 0
\,\,\, \mbox{for} \,\,\, t \to \infty,
\label{eq:balance-eq-T}
\end{equation}
since $\lim_{t \to \infty} \langle \dot{U}(t) \rangle = 0$, 
where $\eta_{\rm ss} \equiv \lim_{t \to \infty} \eta(t)$ and 
${\cal D}_{\rm ss} \equiv \lim_{t \to \infty} {\cal D}(t)$.
Thus, the steady-state kinetic temperature $T_{\rm ss} \equiv \lim_{t \to \infty} T(t)$
is determined by the balance between the viscous heating 
($2\dot{\gamma}^{2} \eta_{\rm ss}/(Dn)$) and the collisional cooling
($4\zeta {\cal D}_{\rm ss}/(Dn)$).

Let us consider an implication of (\ref{eq:balance-eq-T}).
As noted below (\ref{eq:canonical-distribution}), the steady-state averages
such as $\eta_{\rm ss}$ and ${\cal D}_{\rm ss}$ are independent of the
choice of the initial inverse temperature $\beta$, and depend only on the
``thermodynamic" parameters $(N, V, \dot{\gamma})$ and on the
system-specific parameters characterizing elastic repulsive force $f(x)$ in (\ref{eq:F-el-def})
and the viscous function $\zeta(x)$ in (\ref{eq:F-vis-def}), the latter being assumed to be
constant $\zeta(x) = \zeta$ here.
In the following discussion, $N$, $V$, and the parameters specifying $f(x)$
play no significant role, and the dependence of the steady-state averages on them 
shall be suppressed.
We shall also introduce dimensionless quantities, to be distinguished with tilde
such as $\tilde{\dot{\gamma}}$ and $\tilde{\zeta}$, 
which are necessary to properly characterize
the degree of nonequilibrium of the steady state. 
This can be done in terms of $d$, $m$, and $T(t)$ as in Ref.~\citen{Santos04},
or in terms of $d$, $m$, and the spring constant $k$ assuming a linear spring model for $f(x)$.
In any case, one obtains from (\ref{eq:balance-eq-T})
\begin{equation}
\tilde{\dot{\gamma}}^{2} = 
\frac{2 \tilde{\cal D}_{\rm ss}(\tilde{\dot{\gamma}},\tilde{\zeta})}{\tilde{\eta}_{\rm ss}(\tilde{\dot{\gamma}},\tilde{\zeta})} \,
\tilde{\zeta},
\label{eq:balance-eq-T-2}
\end{equation}
which is an analogue of the relation derived in 
Ref.~\citen{Santos04} based on the fluctuating hydrodynamics.
The solution to this equation gives $\tilde{\dot{\gamma}}(\tilde{\zeta})$ that depends only on $\tilde{\zeta}$.
This means that the dimensionless shear rate $\tilde{\dot{\gamma}}$ cannot be made small by
solely controlling the shear rate $\dot{\gamma}$,
and to realize the small degree of nonequilibrium which is characterized by small $\tilde{\dot{\gamma}}$ and $\tilde{\zeta}$,
one has to consider the quasielastic limit $\tilde{\zeta} \ll 1$~\cite{Santos04}.

In the main text, we are interested not only in the steady state which is characterized only by $\tilde{\zeta}$ in the
above sense, but also in the transient dynamics occurring before the steady state is reached.
In the transient regime, the above argument does not hold, and the dynamics there depends on the
shear rate $\dot{\gamma}$ as well.
Thus, one has to take into account both $\dot{\gamma}$ and $\zeta$ in characterizing the transient dynamics.


\begin{thebibliography}{10}
\bibitem{comment-mixing}
The system is said to exhibit mixing if arbitrarily chosen phase variables, say
$A$ and $B$, become uncorrelated at long times, i.e., $\lim_{t \to \infty}
\langle A(t) B(0) \rangle = \langle A(t) \rangle \langle B(0) \rangle$.
\end{thebibliography}
\end{document}